\documentclass[11pt]{article}

\usepackage{lscape,epsfig,cite,texnames,verbatim}
\usepackage{amsmath,amssymb}
\usepackage{times}
\usepackage[ps2pdf,bookmarks=true,bookmarksnumbered,bookmarksopen]{hyperref}

\newcommand{\Ht}{\mathfrak{H}}

\title{General Relativity and Quantum Mechanics: \\Towards a Generalization of the Lambert $W$ Function}
\author{Tony~C.~Scott\textsuperscript{1} and
  Robert Mann\textsuperscript{2} \\[0.5cm]
\footnotesize
    \textsuperscript{1} 
    Institut f\"{u}r Physikalische Chemie, RWTH Aachen, 52056 Aachen, Germany\\
\footnotesize
    Institut f\"{u}r Organische Chemie, Fachbereich Chemie,
    Universit\"at Duisburg-Essen, 45117 Essen, Germany\\
\footnotesize
    Zentralinstitut f\"{u}r Angewandte Mathematik,
    Forschungszentrum J\"{u}lich GmbH, 52425 J\"{u}lich,
    Germany\\
\footnotesize
    email: scott@pc.rwth-aachen.de\\[0.5cm]
\footnotesize
    \textsuperscript{2}  Department of Physics,
    Professor of Physics and Applied Mathematics, University of Waterloo,
    Ontario, Canada N2L 3G1 \\
\footnotesize
    Perimeter Institute for Theoretical Physics, 31
    Caroline Street North, Waterloo, Ontario, Canada N2L 2Y5\\
\footnotesize
    email: {mann@avatar.uwaterloo.ca , rmann@perimeterinstitute.ca}\\[0.5cm]
}
\begin{document}

\maketitle

\begin{abstract}
\par
\noindent
Herein, we present a canonical form for a natural and necessary
generalization of the Lambert $W$ function, natural in that it
requires minimal mathematical definitions for this generalization,
and necessary in that it provides a means of expressing
solutions to a number of physical problems of fundamental nature.
In particular, this generalization expresses the exact solutions
for general-relativistic self-gravitating 2-body and 3-body systems
in one spatial and one time dimension. It also
expresses the solution to a previously unknown mathematical link
between the linear gravity problem and the quantum mechanical
Schr\"{o}dinger wave equation.\\
\vspace*{2ex}
\par
\noindent
{\bf AMS Numbers:} 33E30, 01-01, 01-02\\
Also related to: 70B05, 81Q05, 83C47, 11A99\\

\end{abstract}

\section{Introduction}

The Lambert $W$ function satisfying $W(t) e^{W(t)} = t$ was first introduced by Johann Heinrich Lambert (1728-1777), a contemporary of Euler. Though it
is more than two hundred years old, its importance and universal application
was only realized in the last decade of the 20th century. With the combined efforts of Gonnet and others, the $W$ function has become a tool.
Ironically, it had been injected into the Maple Computer Algebra system by
Gaston Gonnet over many objections because it did not appear as a
``standard'' special function known in literature (\hbox{e.g.}\ {} see
\cite
{AbrSte70a,arfken,Luke}) though it was useful as a means of expressing solutions
to transcendental algebraic equations. A presentation of the work of Scott
\emph{et al.}\cite{Scott1} in 1992 showed that it expressed an exact
solution to a fundamental problem in quantum mechanics.

This encouraged Corless \emph{et al.} to make a literature search of the $W$ function to find
that it had been ``invented'' and ``re-invented'' at
various moments in history. Its applications were numerous\cite{lambert}.
For example, the $W$ function has appeared in electrostatics, statistical
mechanics (\hbox{e.g.}\ {} \cite{statmach}), general relativity,
inflationary cosmology (\hbox{e.g.}\ {} \cite{amjad}), radiative transfer,
Wien's Displacement Law of blackbody radiation (\hbox{e.g.}\ {} \cite{wien}%
), quantum chromodynamics, combinatorial number theory, fuel consumption and
population growth (see \hbox{e.g.}\ {} \cite{generalapplications} and
references herein) etc. Within the past decade, the Lambert $W$ function has
been embedded in other computer algebra systems, multiplying the
applications of this function and also increasing its awareness and thus
vindicating its appearance within a Mathematical software system.

More recently, the Lambert $W$ function has also appeared in the ``lineal''
gravity two-body problem\cite{mann} as a solution to the Einstein Field
equations in $(1+1)$ dimensions. The
Lambert W function appears as a solution for the case when the two-bodies
have exactly the same mass. However, the case of unequal masses required a
\emph{generalization} of Lambert's function\cite[eq.(81)]{mann}. Subsequent
discussion brought the realization that this generalization for unequal
masses had a one-to-one relationship with the problem of unequal charges for
the quantum mechanical problem of Scott \emph{et al.}

This realization fueled the impetus for investigation into a proper
generalization, the focus of this article. Moreover, it became clear that
something vital about the $W$ function had been \emph{missed}. The
information and awareness of the literature on Lambert's function is still
fragmentary. For example,{} long before D.E.G. Hare\cite[(a)]{lambert}
extended the definition of the $W$ function into the complex plane, such an
analysis had already been done indirectly by Byers Brown\cite{byers1,byers2}
not only for the standard $W$ function but also for its generalization
discussed herein. We see that no matter how exhaustive a literature search
is made, it cannot address or cover all the aspects of a function over two
hundred years old!

The goal of the present work is to examine this generalization while
clarifying a number of issues with regards to the unfortunate
``fragmentation'' of information and awareness of Lambert's function. Of
course, one can define generalizations in myriad ways. Thus, we seek a
generalization that is ``natural'' \hbox{i.e.}\ {}

\begin{enumerate}
\item It is economical in that minimizes the need for new mathematical
definitions.

\item It has applications in nature. Better still, it is ubiquitous to
nature, not unlike the standard Lambert $W$ function itself.

\item It expresses solutions to a broad range of mathematical problems.

\item Its capacity for further generalization and its reduction and
correspondence to the standard Lambert $W$ function is transparent.
\end{enumerate}
Such a function satisfying these criteria is clearly a fundamental
mathematical structure worthy of consideration in the Mathematical/Physical
literature.

In this work, we present a canonical form for a generalization of the $W$ function which
satisfies this criteria. This is done as follows. First,
we then re-examine the quantum mechanical problem of Scott \emph{et al.} whose solution also
expresses the solution to the linear gravity problem of Mann and Ohta\cite%
{mann}. We solve the case of unequal charges (unequal masses for the
gravitational problem) intuitively. The impetus is partially derived from
the notion of P.A.M. Dirac that a sound mathematical structure has a
potential basis in reality (and the converse might just be true!).

Next we find that the generalization fits into a \emph{tetration} framework (or iterative exponentiation), and requires only a \emph{nesting} of the definitions of the Lambert W function, thus satisfying the first requirement.
Next, we seek solutions to the gravitational three-body problem in $(1+1)$
dimensions and find that the first generalization can be naturally extended
further. Finally, it is found that the end result expresses solutions to a
huge class of \emph{delayed} differential equations. It is also helpful in
expressing the solution to the (three-dimensional) hydrogen molecular ion %
\hbox{i.e.}\ {} the quantum-mechanical three-body problem for the case of
clamped nuclei. Concluding comments are made at the end.

\section{One-Dimensional Quantum Problem}
\label{sec:prel} The one-dimensional version of the hydrogen molecular ion H$%
_{2}^{+}$\cite{herrick,byers1,byers2} is given by the double Dirac delta
function model:
\begin{equation}
-\frac{1}{2}\frac{\partial ^{2}\psi }{\partial x^{2}}-q[\delta (x)+\lambda
\delta (x-R)]\psi =E(\lambda )\psi  \label{eq:doublewell}
\end{equation}%
where $Z_{A}=q$ and $Z_{B}=\lambda ~q$. The ansatz for the solution has been
known since the work of Frost\cite{frost}:
\begin{equation}
\psi ~=~Ae^{-d|x|}+Be^{-d|x-R|}
\end{equation}%
where $0<R<\infty $. All quantities are real. Matching of $\psi $ at the
peaks of the Dirac delta functions at $x=0,R$ yields:
\begin{equation}
\left|
\begin{array}{cc}
q-d & qe^{-dR} \\
q\lambda e^{-dR} & q\lambda -d%
\end{array}%
\right| =0  \label{eq:mat}
\end{equation}%
and the energies are given by $E_{\pm }=-d_{\pm }^{2}/2$ where $d_{\pm }$ is
governed by the secular determinant of eq.~(\ref{eq:mat}) when it is set to equal zero (\cite[eq.(17)]{Scott1} for $q=1$):
\begin{equation}
d_{\pm }(\lambda )~=~{\textstyle\frac{1}{2}}q(\lambda +1)\pm {\textstyle%
\frac{1}{2}}\left\{ q^{2}(1+\lambda )^{2}-4\,\lambda q^{2}[1-e^{-2d_{\pm
}(\lambda )R}]\right\} ^{1/2}  \label{eq:pseudo}
\end{equation}%
When $\lambda =1$, the pseudo-quadratic in (\ref{eq:pseudo}) reduces to:
\begin{equation}
d_{\pm }=q[1\pm e^{-d_{\pm }R}]  \label{eq:lam}
\end{equation}%
Although, the above has been known for more about half a century, it was not
until the work Scott \emph{et al.}\cite{Scott1} that the solution for $%
d_{\pm }$ was exactly found to be:
\begin{equation}
d_{\pm }=q~+~W(0,\pm qRe^{-qR})/R  \label{eq:sW}
\end{equation}%
where $\pm $ represent respectively the symmetric or \emph{gerade} solution
and the anti-symmetric or \emph{ungerade} solution. The first argument of
the $W$ function, being zero, reminds us that Lambert's function has an
infinite number of branches and that we are selecting the principal branch.

The anti-symmetric case, $d_{-}$ is interesting because it appears to go to
zero as $R\rightarrow 1$ for \linebreak $q=1$ \footnote{For simplicity, we set the charge $q=1$ for the rest of this work.}, in other words the energy goes to zero
and the corresponding eigenstate appears to go into the continuum. However,
for $R<0$, $W(-1,-Re^{-R})$ which has an order 2 branch at $R=1$ yields a
real number. Given the analysis in the complex plane of the energy
eigenstates\cite{byers1,byers2}, we can see there was already awareness of
more than one branch for the solution as far back as the 1970s by
mathematical physicists well versed in the mathematical framework of linear
molecules.

So far, the Lambert $W$ function could only express the solution for the
case of equal charges. We now examine the general case of unequal charges.
The pseudo-quadratic of eq.~(\ref{eq:pseudo}) seems complicated until we
rewrite it in a very simple form \textit{(for $q=1$)}:
\begin{equation}
e^{-2xR}~=~\frac{(1-x)~(\lambda -x)}{\lambda }\quad \mbox{where}\quad
x=d_{\pm }~.  \label{eq:Wgen}
\end{equation}%
The above encapsulates both \textit{``gerade'' }and\textit{\ ``ungerade''}
solutions\footnote{%
Symmetry is lost when $\lambda \neq 1$ and the terms ``gerade'' and
``ungerade'' no longer have the same meaning.}. It also represents a
canonical form for a whole class of transcendental algebraic equations. The
right side of (\ref{eq:P2}) is a quadratic polynomial in $x$ only while the
left-hand side is a function of $x$ and $R$. It must be emphasized that $R$
is a constant in the range $[0,\infty )$ consequently allowing infinite
choices for $R$. The left-hand side of (\ref{eq:Wgen}) is a whole parameter
family of curves in $x$ while the right-hand side represents only one curve
in $x$. Therefore, eq.~(\ref{eq:Wgen}) is a canonical form for an \emph{%
implicit} equation for $x$. Note that when $\lambda =1$, we have a double
root for this polynomial and both sides of (\ref{eq:Wgen}) \emph{factors}
into two possible cases where the solutions are given by eq.~(\ref{eq:sW})
for $q=1$. The problem in linear gravity\cite{mann} namely eq.~$(82)$ of
ref. \cite{mann}:
\begin{equation*}
y^{2}=a^{2}+\left( x^{2}-a^{2}\right) \exp \left( 2x\right) \exp \left(
-2y\right)
\end{equation*}%
relates exactly to eq.~(\ref{eq:Wgen}) by the following transformation:
\begin{equation}
\lambda =\frac{2x}{x+a}-1\text{ \ where \ \ \ }R=-\left( x+a\right) \text{ \
\ and \ }d=\frac{x-y}{x+a}
\end{equation}%
when $q=1$.
The case $\lambda \neq 1$ represents the generalization we seek. Thus, we
seek a solution to:
\begin{equation}
e^{-2xR}~=~a_{o}\,b_{o}(x-r_{1})(x-r_{2})  \label{eq:P2}
\end{equation}%
where in relation to the above problems $\left\{ r_{1},r_{2}\right\}
=\left\{ 1,\lambda \right\} $ are the real roots of a quadratic polynomial
and where $\left\{ a_{o},b_{o}\right\} =\left\{ 1,1/\lambda \right\} $.
However, we treat these parameters generally while making all necessary
assumptions to ensure a real solution. Since a quadratic is merely a product
of first order polynomials, this guides us intuitively to consider the
following. We assume there exists a value $y$ such that:
\begin{eqnarray}
e^{-Rxy} &=&a_{o}\,(x-r_{1})  \label{eq:1} \\
e^{-Rx(2-y)} &=&b_{o}\,(x-r_{2})  \label{eq:2}
\end{eqnarray}%
Multiplication of the left sides and right sides of eqs.(\ref{eq:1}) and (%
\ref{eq:2}) yields (\ref{eq:P2}) the equation we desire to solve. However,
individually eqs.(\ref{eq:1}) and (\ref{eq:2}) can be solved using the
standard Lambert $W$ function:
\begin{eqnarray}
a_{o}(x_{1}-r_{1}) &=&a_{o}\,\frac{W(yRe^{-r_{1}yR}/a_{o})}{yR}
\label{eq:s1} \\
b_{o}(x_{2}-r_{2}) &=&b_{o}\,\frac{W((2-y)Re^{-r_{2}(2-y)R}/b_{o})}{(2-y)R}
\label{eq:s2}
\end{eqnarray}%
Letting $y~=~1+\epsilon $, we seek $y$ such that $x_{1}~=~x_{2}$. Thus, the
``separation'' parameter is governed by:
\begin{equation}
(r_{1}-r_{2})=\frac{W((1-\epsilon )Re^{-r_{2}(1-\epsilon )R}/b_{o})}{%
(1-\epsilon )R}-\,\frac{W(\left( 1+\epsilon \right) Re^{-r_{1}\left(
1+\epsilon \right) R}/a_{o})}{\left( 1+\epsilon \right) R}  \label{eq:sep}
\end{equation}%
Substituting eqs.(\ref{eq:s1}) and (\ref{eq:s2}) into (\ref{eq:P2})
\begin{equation}
e^{-2xR}~=~a_{o}b_{o}\,\frac{W\left( (1+\epsilon )Re^{-r_{1}(1+\epsilon
)R}/a_{o}\right) \,W\left( (1-\epsilon )Re^{-r_{2}(1-\epsilon
)R}/b_{o}\right) }{(1+\epsilon )(1-\epsilon )R^{2}}  \label{eq:sol}
\end{equation}%
and taking logarithms on both sides of (\ref{eq:sol}) allows us to isolate
an expression for $x$ 
subject to the constraint that $y$ is such that $x_{1}=x_{2}$. Looking at
``canonical'' forms of eqs.(\ref{eq:s1}) and (\ref{eq:s2}) in comparison
with (\ref{eq:sol}) makes us infer the \emph{generalized} Lambert W function
as
\begin{equation}
{\Omega }_{2}~=~{\Omega }_{2}(a_{o},b_{o},r_{1},r_{2},R)~=~W(z_{1})~W(z_{2})
\label{eq:W2}
\end{equation}%
where
\begin{eqnarray}
z_{1} &=&(1+\epsilon )Re^{-r_{1}(1+\epsilon )R}/a_{o}  \notag \\
z_{2} &=&(1-\epsilon )Re^{-r_{2}(1-\epsilon )R}/b_{o}  \notag
\end{eqnarray}%
and where $\epsilon =\epsilon (a_{o},b_{o},r_{1},r_{2},R)$. The above is a
product of standard Lambert $W$ functions in the same fashion a quadratic
polynomial is the product of first order polynomials. When $r_{1}=r_{2}$, it
is clear that $\epsilon =0$ (or when an analytical expression for $\epsilon $
is possible) and we recover the solution in terms of the standard Lambert $W$
function. The subscript $2$ on $\Omega $ reminds us that the right side is a
second order polynomial. In general, $y$ or for that matter $\epsilon $ will
be referred to as a \emph{separation} parameter, which allows the
generalized function shown here to be decoupled as a product of standard
Lambert $W$ functions. Note that when $a_{o}=b_{o}=1$, we recover the
symmetric (gerade) solution and with $a_{o}=b_{o}=-1$, the anti-symmetric
(ungerade) solution of eq.~(\ref{eq:sW}).

In view of eq.~(\ref{eq:sep}), one realizes that the {\it separation
parameter} $\epsilon $ is itself governed by a transcendental equation which
looks even more complicated than the original transcendental algebraic
equation of eq.~(\ref{eq:Wgen}). The critic then asks: how are we further
ahead? Part of the answer lies in considering an important aspect about
generalizations \hbox{i.e.}\ {} whether or not they have a capacity to
collapse into special cases other than the original function from which the
generalization was inferred or collapse into previously unknown solutions.
Let $a_{o},b_{o},r_{1},r_{2}$ be the values
as quoted below (\ref{eq:P2}) and let us make a series expansion of $%
x_{1}-x_{2}$ in the parameter $\lambda $:
\begin{eqnarray}
x_{1}-x_{2}~ &\approx &~{\frac{1+{e^{W\left( R\left( 1+\epsilon \right) {%
e^{-R\left( 1+\epsilon \right) }}\right) }}{e^{R\left( 1+\epsilon \right) }}%
}{{e^{\mathit{W}\left( R\left( 1+\epsilon \right) {e^{-R\left( 1+\epsilon
\right) }}\right) }}{e^{R\left( 1+\epsilon \right) }}}}-2\,\lambda
-2\,R\left( -1+\epsilon \right) {\ \lambda }^{2}\quad  \notag \\
&&-4\,{R}^{2}\left( -1+\epsilon \right) ^{2}{\lambda }^{3}-{\ \frac{28}{3}}\,%
{R}^{3}\left( -1+\epsilon \right) ^{3}{\lambda }^{4}-24\,{R}^{4}\left(
-1+\epsilon \right) ^{4}{\lambda }^{5}+O\left( {\ \lambda }^{6}\right) \quad
~
\end{eqnarray}%
By inspection, we can see that $\epsilon =1$ eliminates all terms of order $%
\lambda $ greater than 1 and consequently:
\begin{equation*}
x_{1}-x_{2}~=~\frac{1+e^{W(2Re^{-2R})}e^{2R}}{e^{W(2Re^{-2R})}e^{2R}}%
-2\,\lambda
\end{equation*}%
Solving for $\lambda $ such that the difference $x_{1}-x_{2}$ is zero,
yields:
\begin{equation}
\lambda ~=~\frac{1}{2}+\frac{W(2Re^{-2R})}{4R}  \label{eq:lambda}
\end{equation}%
In this case, the solution for $x$ is found to be:
\begin{equation}
x~=~-\frac{1}{2R}\ln \left( \frac{W(2Re^{-2R})}{2\,R}\right)
\label{eq:xspec}
\end{equation}%
Thus for $\lambda $ satisfying (\ref{eq:lambda}) and $\epsilon \rightarrow 1$%
, we have a previously unknown solution to eq.~(\ref{eq:Wgen}) in terms of
the standard Lambert $W$ function. Granted $\epsilon \rightarrow 1$
represents a limiting extreme: this solution has been vindicated by
numerical and analytical demonstrations using computer algebra. Thus, we
have a previously unknown particular solution for the case of unequal
charges (or unequal masses in the linear gravity problem) but for a peculiar
value of $\lambda $ dependent upon $R$ which admittedly is not physically
useful since the physical charges $Z_{A}$ and $Z_{B}$ are constants
independent of $R$. \ This bears a striking resemblance of the results of
Demkov who found analytical solutions for the three-dimensional hydrogen
molecular ion H$_{2}^{+}$ but for a particular choice of charges (again not
physically useful for the same reasons) which reduced to Whittaker functions %
\hbox{i.e.}\ {} the type of solutions found for the hydrogen atom\cite%
{demkov}.

At any rate, this demonstration shows that the generalization we inferred is
not impotent but can also collapse into simpler special functions for
special cases.  There are other such cases. Since eq.(\ref{eq:sep}) has the form
$r_1 - r_2 = f(r_2 ) - g(r_1)$, one particular obvious solution is
simply $r_1=f(r_2)$ and $r_2=g(r_1)$. Decoupling this particular solution
with respect to $r_1$ and $r_2$ yields equations governing the latter:
\begin{eqnarray}
r_1 & = & \frac{W ( R(1-\epsilon)
\exp ( \frac{(\epsilon-1)
W\left( (1+\epsilon)
e^{-r_1 R (1+\epsilon )}/a_o \right)}{1+\epsilon})/b_o
)}{R(1-\epsilon )} \label{eq:param} \\
r_2 & = &  \frac{W ( R(1+\epsilon)
\exp ( \frac{(1-\epsilon)
W\left( (1-\epsilon)
e^{-r_2 R (1-\epsilon )}/b_o \right)}{1-\epsilon})/a_o
)}{R(1+\epsilon )}  \nonumber
\end{eqnarray}
At this stage, we can change our context: rather than look for solutions
of eq.~(\ref{eq:P2}) for arbitrary choices of the
parameters $R$, $a_o$, $b_o$, $r_1$ and $r_2$
and be burdened with the analytical determination of $\epsilon$:
we instead use $\epsilon$ as a {\em common} parameter to $r_1$ and
$r_2$ allowing us to find which values of $r_1$ and $r_2$
satisfy eq.~(\ref{eq:P2}) for a given choice of the remaining
parameters $a_o$ $b_o$ and $R$.  Moreover, we also find the following
previously unknown {\em exact} solutions to (\ref{eq:P2}):
\begin{equation}
r_1 = \frac{1}{b_o}, \quad
a_o = \frac{e^{\frac{-2R (1+b_o r_2 )}{b_o}}}{r_2} , \quad
x = \frac{1+b_o r_2 }{b_o} , \quad \epsilon \rightarrow 1 \, .
\label{eq:spec1}
\end{equation}
where $r_2$ and $b_o$ are arbitrary real numbers and also
\begin{equation}
r_2 = \frac{1}{a_o}, \quad
a_o = -\frac{2R}{2 r_1 R + \ln( r_1 b_o )} , \quad
x = -\frac{1}{2~R} ~ \ln(r_1 b_o )  \, ,
\label{eq:spec2}
\end{equation}
where $r_1 $ and $b_o$ are arbitrary real numbers.
Eq.~(\ref{eq:spec1}) is in {\em closed} form (elementary functions)
and (\ref{eq:spec2}) is
in terms of the Lambert W function which bears some resemblance
to eq.~(\ref{eq:xspec}) and could be called another ``Demkov''-type
solution.
This is one of the utilities of the ``separation'' parameter $\epsilon$:
eqs.~(\ref{eq:sep}) and especially (\ref{eq:param}) look more
complicated than the original equation
(\ref{eq:P2}), they are however useful in finding
its particular solutions\footnote{%
We anticipate the existence of other cases when $\epsilon $ can be solved in
closed form but of course, in general, this is not the case. Indeed, $%
\epsilon $ might not even exist which is why the generalization is needed.}.
The parameter $\epsilon$ also illustrates the reduction
from the proposed generalized function to the simpler
(standard) Lambert $W$ function
in a transparent manner. In the next section, we properly define the
function ${\Omega }_{2}$ which we derived intuitively.

\section{Iterated Exponentiation} \label{sec:martinez} 
Infinitely iterated exponentiation or {\it tetration} is defined as the limit,
\begin{equation}
\lim_{n\rightarrow \infty }\, ^{\alpha} n ~=~\alpha ^{{\alpha }^{{\alpha }^{{%
\cdot }^{{\cdot ^{\cdot }}}}}}~\equiv ~\mathfrak{H}(\alpha )
\end{equation}%
This function can be written compactly as:
\begin{equation}
\Ht (\alpha )= e^{-W(-\ln \alpha )}  \qquad \alpha \in \mathbb{C} \label{eq:H}
\end{equation}%
The product on $n$ $W$ functions is given by:
\begin{equation*}
W(z_{1})W(z_{2})\ldots W(z_{n})=(z_{1}z_{2}\ldots z_{n})e^{-[W(z_{1})+W(z_{2})+\ldots +W(z_{n})]}
\end{equation*}%
following from the defining relation $z=W(z) e^{W(z)}.$ The Lambert $W$
function is governed by the addition law\cite{addition}:
\begin{equation}
W(a)+W(b)~=~W(ab\,\left[ 1/W(a)+1/W(b)\right] )
\end{equation}%
for $\Re (a)$ or $\Re (b)>0$. By induction, the product is then:
\begin{eqnarray}
W(z_{1})\,W(z_{2})\ldots W(z_{n}) &=&(z_{1}z_{2}\ldots z_{n})~e^{-W(f_{n})}  \notag \\
&=&(z_{1}z_{2}\dots z_{n})~W(f_{n})/f_{n}
\end{eqnarray}%
where $f_{n}$ stands for an expression involving $W $ functions of $%
z_{1}$, $z_{2}\ldots z_{n}$ that follows from the addition law. \hbox{E.g.}\
{} for $n=2$ and $n=3$,
\begin{eqnarray}
f_{2} &=&z_{1}z_{2}~(1/W(z_{1})+1/W(z_{2}))  \notag \\
f_{3} &=&f_{2}z_{3}~(1/W(f_{2})+1/W(z_{3}))  \notag
\end{eqnarray}%
Thus,
\begin{eqnarray}
W(z_{1})\,W(z_{2})
&=&z_{1}z_{2}\,W(f_{2})/(z_{1}z_{2}(1/W(z_{1})~+~1/W(z_{2})))  \notag \\
&=&W(f_{2})/(1/W(z_{1})~+~1/W(z_{2})).
\end{eqnarray}%
In the context of the tetration function $\Ht (\alpha )$ of (\ref%
{eq:H}), the \textit{product of Lambert $W$ functions} is really just:
\begin{equation}
{\Omega }_{n}(z_{1},z_{2},\ldots ,z_{n})=W(z_{1})W(z_{2})\dots
W(z_{n})=(z_{1}z_{2}\dots z_{n})\mathfrak{H}(e^{-f_{n}})  \label{eq:Wn}
\end{equation}%
a multiple of the continued tetration function. To reiterate, for eqs.(\ref%
{eq:s1}) and (\ref{eq:s2}) when $x=x_{1}=x_{2}$, the solution to eq.~(\ref%
{eq:P2}) is given by a product of $W$ functions, as one substitutes $%
x_{j}$ into $(x-r_{j})$ for $j=1,2$. Hence, the solution is in terms of one $%
W$ function, albeit nested. In this framework, thanks to the addition
law for the $W$ function, no generalization in the sense of additional
mathematical definitions is needed: it is simply one single $W$ function evaluated at a point involving other $W$ functions.

Furthermore, if we use eq.~(\ref{eq:Wn}) for $n=2$ and combine with the
earlier result in (\ref{eq:W2}), we obtain a simple yet general relation
between the separation parameter $\epsilon $ and $x$:
\begin{equation}
(r_{2}-r_{1})\epsilon ~=~\frac{W(f_{2})}{R}~+~(r_{1}+r_{2})-2~x
\end{equation}%
which is consistent with eq.~(\ref{eq:sol}) in view of the addition theorem
for $W$ functions. We can already identify regimes. For example,\ for real
positive roots $r_{1}$ and $r_{2}$ 
the term $W(f_{2})$ is bounded in many cases and consequently\footnote{%
Of course, we realize that it is much easier to solve the transcendental
equations numerically; the goal of this exercise is to define and justify
our generalization of the $W$ function.} $\lim_{R\rightarrow \infty
}\epsilon =\pm 1$.

Although we initially assumed $\Re ~{z_{n}}$ is not a negative number,
injecting of numbers into eq.~(\ref{eq:Wn}) show that this identity formula
is very robust after all if one considers different $W$ functions on
different branches. For example,\ {} (\ref{eq:Wn}) holds if $x_{1}=1+2i$, $%
x_{2}=-1$ for $W(x_{2})=W(-1,x_{2})$ and is complex-valued.

\section{Rational Polynomials}

In the previous sections, we considered the right side of the transcendental
equations to be polynomials, but one can also consider a rational
polynomial. Let us consider the equation where the right side is a ratio of
first order polynomials:
\begin{equation}
e^{-2 R x } ~=~ \frac{a_o (x-r_1 )}{b_o (x-r_2 )}  \label{eq:1m1}
\end{equation}
In parallel to what we did before, we can consider:
\begin{eqnarray}
e^{-R x y} &= & a_o \, (x-r_1 )  \label{eq:3} \\
e^{-R x (2-y)} &= & \frac{1}{b_o \, (x-r_2 )}  \label{eq:4}
\end{eqnarray}
Eq.(\ref{eq:4}) can also be solved in terms of the standard Lambert $W$
function \hbox{i.e.}\ {} once the inverse is taken on both sides of (\ref%
{eq:4}) \hbox{i.e.}\ {} $(2-y) \rightarrow (y-2) $ in eq.~(\ref{eq:2}).
The generalization is thus:
\begin{equation}
{\ \Omega }_{1,-1} ~=~ W \left( (1 + \epsilon) R e^{-r_1 (1 + \epsilon )
R}/a_o \right) \, W \left( -(1 - \epsilon) R e^{+r_2 (1- \epsilon ) R}/b_o
\right)
\end{equation}
where the subscripts $1,-1$ denote respectively the polynomial degrees in (%
\ref{eq:1m1}). This also fits into the tetration framework of section \ref%
{sec:martinez}, the only change being done to the argument $x_2$ and the
right side of (\ref{eq:1m1}) can be generalized to a rational polynomial of
higher order as both numerator and denominator can be ``separated'' in a
manner demonstrated in eqs.(\ref{eq:3}) and (\ref{eq:4}) to yield
generalized $\Omega$ functions that can be further nested together using the
addition theorem. The next section presents applications of this type of
problem.

\section{Three-Body Linear Gravitational Motion}

The solution of the three-body one $(1+1)$ dimensions via dilation theory
requires solving for $V$ which is governed by the following equation%
\cite[eq.(32)]{mann2}:
\begin{equation}
(V-m_{1})~(V-m_{2})~(V-m_{3})\qquad \qquad \qquad \qquad \qquad \qquad
\qquad \qquad ~~
\end{equation}%
\begin{eqnarray}
\qquad &=&m_{1}~m_{2}~(V-s_{1}~s_{2}~m_{3})~\exp (K~V~R~|\sin (q)|)  \notag
\\
&+&m_{1}~m_{3}(V+sq~s_{2}~m_{2})~\exp (K~V~R~|\sin \left( q+\frac{\pi }{3}%
\right) |)  \notag \\
&+&m_{2}~m_{3}~(V-sq~s_{1}~m_{1})~\exp (K~V~R~|\sin \left( q-\frac{\pi }{3}%
\right) |)  \notag \\
&&
\end{eqnarray}%
where
\begin{equation*}
\begin{array}{lccl}
s_{1} & = &  & \mbox{sign}(\sin (q+\frac{\pi }{3})) \\
s_{2} & = & - & \mbox{sign }(\sin (q-\frac{\pi }{3})) \\
sq & = &  & \mbox{sign }(\sin (q))%
\end{array}%
\end{equation*}%
This problem has no closed form solution. The hard part is in
trying to extricate the exponential terms from the rest of the
expression. At $q=0$ or $\pm \pi /3$, the trigonometric quantities
$s_{i}$ for $i=1,2$ and $sq$ simplify to specific values.
\hbox{E.g.}\ {} for $q=0$ and $V\neq 0$, it is found that:
\begin{equation}
\exp (-2R_{t}~V)~=~\frac{1}{m_{3}(m_{1}+m_{2})}~(V-m_{3})(V-(m_{1}+m_{2}))
\label{eq:m3}
\end{equation}%
where $R_{t}~=~KR\sqrt{3}/4$. As we can see $V$ is governed by exactly the
same type of transcendental equation as eq.~(\ref{eq:P2}) whose solution, as
we have seen, can be expressed in terms of our $\Omega _{2}$ function. In
the case when $m_{3}=m_{1}+m_{2}$, we have a double root and thus a solution
for (\ref{eq:m3}) can be expressed exactly in terms of the standard $W$
function and has two solutions:
\begin{equation}
V~=~m_{3}-\frac{1}{R_{t}}W\left( \pm ~R_{t}~m_{3}e^{R_{t}m_{3}}\right) \quad %
\mbox{when}\quad q=0,~m_{3}=m_{1}+m_{2}
\end{equation}%
Very similar solutions also exists for $q=\pi /3$ with $m_{1}=m_{2}+m_{3}$
and for $q=-\pi /3$ and $q=2\pi /3$ with $m_{2}=m_{1}+m_{3}$. This
demonstrates that at every $\pi /3$, solutions in terms of our generalized $%
\Omega $ functions exist. At $q=\pi /6$ and equal masses \hbox{i.e.}\ {} $%
m_{1}=m_{2}=m_{3}$ we arrive at two possible equations depending on the
outcome of factorization.  One equation is:
\begin{equation}
\exp (cV) =-\frac{(V-m)}{m} \, ,  \label{eq:P1}
\end{equation}
and the other equation is:
\begin{equation}
\exp (cV) =-\frac{(V-m)^{2}}{m(V+m)}  \, , \label{eq:P2Q1}
\end{equation}
where $c=KR/2$. Eq.(\ref{eq:P1}) can be solved in terms of the standard $W$
function \hbox{i.e.}\ {}
\begin{equation*}
V~=~m \left( 1 -\frac{W(mc e^{mc})}{mc} \right) \quad
\end{equation*}%
which is in the same form as the range $x_{\text{range}}$ and drift $z_{\text{drift}}$ equations of section 2 and the anti-symmetric solution $d_{-}$ in section 3 with the exact correspondence given by $\eta = - mc$ and $b = v_{\text{x},0}$. However the right side of (\ref{eq:P2Q1}) involves a rational polynomial of the
form $P_{2}(V)/Q_{1}(V)$ and requires $\Omega _{2,-1}$ to express the
solution. If we consider the region $0<q<\pi /3$, we obtain the form:
\begin{equation}
\exp(-c~\sqrt(3)~\cos(q) V) =\frac{P_N (V)}{Q_M (V)}\quad \mbox{where}\quad M,N \rightarrow \infty
\end{equation}%
This rational polynomial can be generated from:
\begin{equation*}
\frac{f~(V-m_3)~(V2 - V*(m_1 + m_2) - m_1*m_2 (f^{2}-1))}{m_3*(m_1*m_2*(f^{2}-1) + m_1*f^{2} + m_2*V)} \rightarrow  \frac{P_N (V)}{Q_M (V)} 
\end{equation*}%
where $f=\exp(c\, \sin (q) V)$ generates the rational polynomials.
In this regime $\sin (q)\approx q$ is small
and $\cos (q)\approx 1$ and thus this treatment is possible. A similar story
applies to other regimes $[\pi /3,2\pi /3]$, $[2\pi /3,\pi ]$ etc \ldots
Thus, we can see that the general solution falls in the following general
form.

\section{General Form}

Thus, the fully generalized form concerns expressing solutions to this
general class of transcendental algebraic equation:
\begin{equation}
e^{\pm k\,x}~=~\frac{P_{N}(x)}{Q_{M}(x)}  \label{eq:general}
\end{equation}%
where $k>0$ is a constant and $P_{N}(x)$ and $Q_{M}(x)$ are polynomials in $%
x $ of respectively orders $N$ and $M$. \ The general solution can be
expressed by $\Omega _{N,M}$ which is formally a product of $N+M$ (standard)
Lambert $W$ functions with $N+M-1$ \ ``separation'' constants.

The standard $W$ function applies for cases when $N=1$ and $M=0$ and
expresses solutions for the case of equal charges for eq.~(\ref%
{eq:doublewell}) or the case of equal masses for the lineal two-body $(1+1)$
gravity problem. Correspondingly, the case $N=2$ and $M=0$ expresses
solutions for the case of unequal charges (or unequal masses for the lineal
two-body gravity problem) and some particular cases of the lineal three-body
gravity problem.\ \ In the limit as $M,N\rightarrow \infty $, this equation
can be used to express solutions of the three-body lineal gravity problem as
we have just shown.

Moreover, the case $N=2$ and $M=0$ (and more generally $N=2$ and $M=1$ )
express the solutions of a significant class of delayed differential
equations \cite[eq.(3)]{sueann}. These arise in a variety of mechanical or
neuro-mechanical (oscillatory) systems in which non-linear feedback plays an
important role.  These have applications in \hbox{e.g.}\ {} models
for physiological systems (medicine)\cite{medicine}.

Recently  Adilet Imambekov and Eugene Demler\cite{bosefermi} considered
a bose-fermi mixture in one dimension.  We note that their Eq.$(2)$:
\[
H = - \sum_i^N \frac{\partial^2}{\partial x_i^2}
+ 2c \sum_{i<j} \delta (x_i - x_j ) \quad c > 0
\]
is in fact the Schr\"{o}dinger wave equation for a {\em linear molecule}.
Their analytical solution as written in their eq.~$(20)$ is indeed a
special case of the general form expressed (\ref{eq:general}).
Thus, we can
also see that this function has fundamental applications in Physics
and can play a fundamental role in Mathematics.

\section{Conclusions}

Thus we have identified a generalization of the Lambert $W$ function or
Omega function as solutions to a large class of {\em transcendental equations} as written in eq.~(\ref{eq:general}). This generalization, denoted
$\Omega _{n}$, satisfies the criteria mentioned earlier in our introduction
by using the {\em analytic} framework of tetration.

We have also shown that the two-body problem in lineal gravity and double
well linear quantum mechanics have the same generalization of $W$, namely $%
\Omega _{2}$ as solutions. The reason why is because the linear gravity
theory via dilaton theory produces a partial differential equation, namely
eq.~$(30)$ of ref. \cite{mann} which can be treated formally as the Schr\"{o}%
dinger wave equation as written in (\ref{eq:doublewell}). This is to be
elaborated elsewhere \cite{mann3}.

Furthermore, this work on the Lambert $W$ function has helped in finding
analytic solutions to the quantum mechanical 3-body problem known as the
hydrogen molecular ion in the case of clamped nuclei of equal charges\cite%
{h2plus}. All this vindicates our proposed generalization of the $W$
function as being of fundamental and physical importance.

Most of the special functions in the known literature (\hbox{e.g.}\ {} \cite%
{AbrSte70a}) are special cases of the hypergeometric functions and/or the
Meijer $G$-function\cite%
{Luke}. The Lambert $W$ function apparently bears no
relationship to these functions and belongs to a class of its own. The
generalization we have presented is a first step in identifying that class.

Although we have inferred a canonical form for a generalization as
expressed by (\ref{eq:general}) and given mathematical and
physical justifications for it, we have yet to clearly identify a
domain and range of applicability or conditions of analytic
continuation. Neither have we formulated Taylor series nor
asymptotic series useful for computation.
Naturally, it is expected that our ``separation'' parameter $\epsilon$ in (%
\ref{eq:sol}) likely has a restricted domain of applicability (its primary
use being to infer the general result starting from the standard $W$
function).

Nonetheless, given that we have fast computational means to transcendental
algebraic equations, the task of obtaining the floating-point attributes in
many cases is trivial today with modern algorithms. Ironically, it has been
our capacity to readily solve these transcendental equations numerically
that has made many take an analytical solution for granted. This may have
been a mistake as we can see that such analytical solutions are ubiquitous
to certain fundamental problems in Physics and Mathematics. In a true sense,
this work is only a beginning.

\addtocontents{toc}{\vspace{0.3cm}}
\addcontentsline{toc}{section}{\numberline {Acknowledgments} \hspace*{2cm} }
\section*{Acknowledgments}

One of us (T.C.S.) would like to thank
Professor Arne L\"{u}chow of the
Institut f\"{u}r Physikalishe Chemie, RWTH-Aachen and Professor Georg Jansen
of the Institut f\"{u}r Organische Chemie of the University of Essen for
their wonderful hospitality and support for allowing this work to be
possible.
We would also like to thank Dirk Andrae of the Theoretical
Chemistry group at the University of Bielefeld (Faculty of Chemistry) and
James Babb and Alexander Dalgarno of the Institute for Theoretical Atomic
and Molecular Physics at the Harvard-Smithsonian Center for Astrophysics,
for helpful discussions.  This work was supported in part by the Natural Sciences and Engineering Research Council of Canada.

\addtocontents{toc}{\vspace{0.3cm}}
\addcontentsline{toc}{section}{\numberline {References} \hspace*{1cm}}

\end{document}